\documentclass{PoS}

\newcommand{\F}{{\cal F}}
\newcommand{\N}{{\cal N}}

\newcommand{\Z}{\mathbb{Z}}
\newcommand{\Ka}{K{\"a}hler }
\renewcommand{\O}{{\cal O}}
\renewcommand{\Re}{{\rm Re}\, }
\renewcommand{\Im}{{\rm Im}\, }

\newcommand{\abs}{|}

\newcommand{\Str}{\textrm{Str}\, }

\newcommand{\ie}{{\em i.e. }}
\newcommand{\via}{{\it via} }

\newcommand{\when}{\mbox{when}}
\renewcommand{\and}{\mbox{and}}

\newcommand{\V}{{\cal V}}

\title{Stringy N = 1 super no-scale models}

\ShortTitle{super no-scale}

\author{\speaker{Costas Kounnas}
\\
        Laboratoire de Physique Th\'eorique,
Ecole Normale Sup\'erieure\thanks{Unit{\'e} mixte  du CNRS et de l'Ecole Normale Sup{\'e}rieure associ\'ee \`a l'Universit\'e Pierre et Marie Curie (Paris 6).\\ \hspace{-0.7cm}\small LPTENS 15/05, CPHT-RR052.1015}, CNRS, 24 rue Lhomond, F--75231 Paris cedex 05, France\\
        E-mail: \email{Costas.Kounnas@lpt.ens.fr}}

\author{Herv\'e Partouche\\
        Centre de Physique Th\'eorique, Ecole polytechnique, CNRS, Universit\'e Paris-Saclay,\\ F--91128 Palaiseau cedex, France\\
        E-mail: \email{herve.partouche@polytechnique.edu}}

\abstract{$\N=1$ no-scale models describe at tree level the spontaneous breaking of supersymmetry at an arbitrary scale $m_{3/2}$, with vanishing vacuum energy. We define $\N=1$ super no-scale models in string theory as being those, which maintain these properties at 1-loop. In other words, in super no-scale models, $m_{3/2}$ is a flat direction of a positive semi-definite 1-loop effective potential. We find explicit examples in heterotic $\Z_2\times \Z_2$ orbifold models, where $\N=1$ is spontaneously broken by a stringy Scherk-Schwarz mechanism, and where the ``decompactification problem'' does not arise.}

\FullConference{18th International Conference From the Planck Scale to the Electroweak Scale \\
		25-29 May 2015\\
		Ioannina, Greece}

\begin{document}


\section{Introduction}

At present, superstring theory is the only framework in which all known interactions, including gravity, are consistently described and unified at the quantum level. Thus, besides questions about the cosmological evolution of the Universe, string theory can address issues in particle physics, and as a first step, the latter can be considered at the classical level in four dimensional Minkowski spacetime. Thirty years of efforts led to several $\N=1$ supersymmetric string models having quasi-realistic spectra, with a net number of three chiral families and containing the standard model interactions at low energy. Among them, are those realized in fermionic construction in heterotic string having an $SO(10)$ gauge symmetry broken at the string level by discrete Wilson lines to the Pati-Salam gauge group, $SU(4)\times SU(2)_L\times SU(2)_R$, which is further broken to the Standard Model gauge group, $SU(3)\times SU(2)_L\times U(1)_Y$, by a usual Higgs mechanism \cite{classi}. However, since all these quasi-realistic models possess $\N=1$ supersymmetry (susy) at tree level, the question of how it is broken and {\em a priori} at a low scale to offer a solution to the gauge hierarchy problem arises. Other questions are then following. Are there conditions insuring perturbation theory to be  defined consistently ?  And, what are the implications of the susy  breaking for the cosmological constant ? The aim of the present work is to provide elements of answers to these problems.

All known supersymmetry breaking mechanisms that avoid the generation of a large cosmological constant, say of order of the string scale, yield effective $\N=1$  no-scale supergravity theories \cite{noscale}. By definition, the latter describe at tree level a spontaneous breaking of supersymmetry at an arbitrary scale $m_{3/2}$, while the cosmological term vanishes. In other words, $m_{3/2}$ is a flat direction of a  positive semi-definite classical potential $\V_{\rm tree}$. Defining $S$ to be the dilaton-axion field and $T_I, U_I$ the \Ka and complex structure moduli characterizing the internal 6-dimensional space (their numbers are in general distinct), one has
\begin{equation}
\langle \V_{\rm tree} \rangle= 0\; ,  \qquad \qquad m^2_{3/2}={\abs w_0\abs^2\over \Im z_1\, \Im z_2\, \Im  z_3}\, , 
\end{equation}
where $w_0$ is the holomorphic superpotential and the choice of scalars $z_1,z_2,z_3$ among the set $\{S,T_I,U_I\}$ is  model-dependent. The superpotential being independent of at least one of the $z_i$'s, the latter is undetermined by the condition $\langle \V_{\rm tree} \rangle= 0$ and $m_{3/2}$ is a flat direction. 

Starting from a  classical string model that is $\N=1$  supersymmetric in Minkowski spacetime, a modification of the superpotential is required for $m_{3/2}$ to be non-trivial. Several mechanism can be invoked :
\begin{description}
\item[\footnotesize $\bullet$] At the level of the supergravity, a non-perturbative mechanism such as gaugino condensation may induce a stabilization of the dilaton \cite{gc}. This yields $z_3=S$ and $\abs w_0(S)\abs^2/\Im S$ equal to ${\Lambda_{\rm np}^3\over M_{\rm s}}$ at the minimum of the potential, where $\Lambda_{\rm np}$ is the generated scale and $M_{\rm s}$ is the string scale $1/\sqrt{\alpha'}, $\footnote{If the dilaton $\Phi$ is split into a vacuum expectation value (VEV) and a fluctuation, $\Phi=\langle\Phi\rangle+\phi$, and the Einstein-Hilbert action is normalized with the Planck mass squared, ${1\over 16\pi G_N}\equiv M_{\rm P}^2/2\equiv e^{-2\langle\Phi\rangle}M_{\rm s}^2/2$, then $\Im S$ is defined to be $e^{-2\phi}$ and its VEV is by construction equal to 1.} while $z_1$ and $z_2$ belong to the set $\{T_I\}$. In this approach, the string predictability is partially lost since the effective susy breaking parameters, which are proportional to the gaugino condensation scale $\Lambda_{\rm np}$, are known up to proportionality coefficients. 
\item[\footnotesize $\bullet$] Perturbative and/or non-perturbative fluxes \cite{Fluxes} along internal directions can be implemented to induce a superpotential responsible for the susy breaking. In some cases, the use of $S$-, $T$- or $U$-dualities between the heterotic, type IIA, type IIB and orientifold theories leads to a partial predictability of the setup \cite{StringDualities}. Different choices of $z_i$'s may be equivalent, while passing  from one picture to another. For instance, they can belong to the sets $\{T_I\}$, $\{U_I\}$, $\{S,T_1,U_1\}$, etc.  
\end{description} 
A particular choice of geometrical fluxes induces a stringy Scherk-Schwarz (SSS) supersymmetry breaking mechanism \cite{SS,SSstring}. It has the advantage to be introduced at the perturbative string level in orbifold compactifications \cite{orbifolds}, or more generally in fermionic constructions \cite{fermionic}. The susy breaking parameters can be  directly  computed from perturbative string amplitudes. For instance, one can implement in heterotic $\Z_2\times Z_2$ fermionic construction the marginal deformations associated to the \Ka and complex structures of the three internal 2-tori, which are parameterized by the complex moduli $T_I,U_I$, $I=1,2,3$ \cite{modulidef,N=0thresh}. In this ``moduli-deformed fermionic construction'', if for simplicity the susy breaking involves a single internal circle of radius $R$, one has 
\begin{equation}
\label{m232}
m_{3/2}^2\propto {M_{\rm s}^2\over \Im S\, \Im T_1\, \Im U_1}= \left({e^{\phi}M_{\rm s}\over R}\right)^2\, .
\end{equation}
Imposing $m_{3/2}$  to be small compared to the string scale, say of order $1-10 \; \rm TeV$,  with $M_{\rm s}=\O(10^{17})$ GeV, one concludes that the internal radius is enormous, $R=\O(10^{13})$. Important consequences for the gauge couplings and the cosmological constant follow.

When $R$ is very large, infinite towers of light Kaluza-Klein states (KK) exist, giving rise in general to large quantum corrections \cite{AntoniadisTeV}
. This fact imposes strong constraints on the model, for perturbation theory to be valid. For instance, in the case of an asymptotically free gauge group factor $G^i$, an almost perfect compensation of the tree level coupling (which is essentially the string coupling $g_{\rm s}=e^{\langle \Phi\rangle}$) and the 1-loop correction must occur, for the  effective gauge coupling at 1-loop to be of order 1. This fine-tuning is a manifestation of the so-called ``decompactification problem''. 

A large value of $R$ also yields a very specific form of the 1-loop effective potential that is proportional to $(n_{\rm F}-n_{\rm B})m_{3/2}^4$, where $n_{\rm F}$ and $n_{\rm B}$ are the numbers of massless fermionic and bosonic degrees of freedom. If, as follows from the fact that the theory is in a spontaneously broken phase of supersymmetry, there is no contribution of order $M_{\rm s}^4$, no term of order $M_{\rm s}^2m_{3/2}^2$ arises either, due to the presence of an infinite tower of light KK states. This is important since such a contribution would otherwise lead to a destabilization of the gauge hierarchy, which requires $\langle m_{3/2}\rangle \ll M_{\rm s}$. Actually, minimizing with respect to $m_{3/2}$, one would obtain either $\langle m_{3/2}\rangle=0$ (no susy breaking) or $\langle m_{3/2}\rangle=\O(M_{\rm s})$ (no hierarchy) \cite{FKZ}. 

However, the cosmological term arising at 1-loop being of order $m_{3/2}^4$, it is still by far too large, compared to the presently observed one, which is of order $10^{-120}$ in Planck units. Thus, we introduce a subclass of models, where the no-scale structure valid at tree level remains true at $\mbox{1-loop}$. These ``stringy super no-scale models'' have a positive semi-definite 1-loop effective potential that admits $m_{3/2}$ as a flat direction \cite{snc},
\begin{equation}
\langle \V_{\mbox{\scriptsize 1-loop}} \rangle= 0\; ,  \qquad \qquad m^2_{3/2}={\abs w_0\abs^2\over \Im z_1\, \Im z_2\, \Im  z_3}\, ,
\end{equation}
provided $m_{3/2}\ll M_{\rm s}$. They are characterized by the condition $n_{\rm F}=n_{\rm B}$.

In Sect. \ref{pbs}, we give a brief overview of the above  issues. We explain in more details the decompactification problem, in the context of effective gauge couplings. Then, we derive the $m_{3/2}^4$ scaling of the 1-loop effective potential arising in string or KK field theory, when susy is spontaneously broken by a (stringy) Scherk-Schwarz mechanism. This behavior will motivate our definition of super no-scale models.

Sect. \ref{d-pb} is a review of a solution to the decompactification problem \cite{N=0thresh}, in the setup of the so-called heterotic string $\Z_2\times\Z_2$ ``moduli-deformed''  fermionic construction. It applies to models where (at least) one of the $\Z_2$ actions is free. However, it is important to mention that this requirement implies the models are not chiral, which means that a fully satisfactory context for describing realistic, 
non-supersymmetric, chiral models still remains to be found. 

Nevertheless, some of the models presented in Sect. \ref{d-pb} admit a super no-scale structure. This is shown in  Sect. \ref{sns}, where the first examples of super no-scale models are presented \cite{snc}.  

Finally, Sect. \ref{cl} summarizes briefly our results. 


\section{On the decompactification and hierarchy problems}
\label{pbs}

Our aim in this section is to present  difficulties encountered in no-scale models realized in string theory, when the susy breaking is implemented \via SSS mechanism. We discuss the 1-loop corrections to the gauge couplings that may be large as the gravitino mass is low, which yields the decompactification problem. We then describe the $m_{3/2}^4$  scaling of the effective potential, which is fine concerning the gauge hierarchy but inadequate for the cosmological constant problem. 


\subsection{Gauge couplings}

We consider the heterotic string in four dimensions. For any gauge group factor $G^i$, the  low energy running gauge coupling takes at 1-loop the general form  \cite{thresholds,universality, FlorakisN=0}
\begin{equation}
\label{running}
{16\pi^2\over g^2_i(\mu)}=k^i{16\pi^2\over g^2_{\rm s}}+b^i\ln {M^2_{\rm s}\over \mu^2}+\Delta^i\, ,
\end{equation}
where the tree level part involves the Kac-Moody level $k^i$ and  the string coupling $g_{\rm s}$. The massless states, with $\beta$-function coefficient $b^i$ yield the logarithmic term, which depends on the energy scale $\mu$. The contribution of the massive modes is denoted $\Delta^i$, the so-called threshold corrections to the gauge coupling. Our aim is to show how the decompactification problem arises in a simple case, namely that of orbifold compactification or moduli deformed fermionic construction, when an internal circle of radius $R$ is large and factorized. 

In these circumstances, $\Delta^i$ takes into account the supermassive string states together with a full tower of charged KK states of masses $m/R$, $m\in \Z$. Except in the $\N=4$ maximally supersymmetric case in which $b^i$ and $\Delta^i$ vanish, the threshold corrections are dominated by the KK states.\footnote{The arguments of this claim are given below Eq. (\ref{kk}), when we compute the effective potential.} For instance, in the $\N=2$ supersymmetric case realized by a $T^2\times K3$ compactification\footnote{The result in universal \ie invariant, wether one considers $K3$ at an orbifold point or not.}, taking one of the radii of $T^2$ to be large, one obtains  
\begin{equation}
\label{D}
\Delta^i=C^iR-b^i \ln R^2+\O\left({1\over R}\right) ,
\end{equation}
where $C^i=Cb^i-C'k^i$, for some positive $C$ and $C'$, which dependent on the second radius of $T^2$~\cite{solving}. Since $C^i=\O(1)$, requiring in Eq. (\ref{running}) the loop correction to be small compared to the tree level contribution amounts to having $g_{\rm s}^2 R<1$. In other words, for perturbation theory to be valid, the string coupling  must  be extremely weak, $g_{\rm s}<\O(10^{-6.5})$. When $C^i>0$, which implies  $G^i$ is not asymptotically free, Eq. (\ref{running}) imposes the running gauge coupling to be  essentially free, $g_i(\mu)=\O(g_{\rm s})$, and $G^i$ describes a hidden gauge group. However, when $C^i<0$, which is the case if $G^i$ is asymptotically free, the very large tree level contribution proportional to $1/g_{\rm s}^2$ must cancel $C^i R$, up to very high accuracy, for the running gauge coupling to be of order 1. Furthermore, this unnatural tuning of $g_{\rm s}$ cannot even be imposed by hand, when several $G^i$-factors coexist, with distinct couplings. 

This phenomenon, known as the ``decompactification problem'', may actually arise each time a submanifold of the internal space is large, in string units, \ie when the internal CFT can be interpreted as a geometrical compactified space. For example, this happens in the supersymmetric compactifications on Calabi-Yau threefolds, in the large volume scenari. In the case we are mostly interested in the present work, $R$ determines as in Eq. (\ref{m232}) the scale $m_{3/2}$ of spontaneous $\N=1$ supersymmetry breaking. Given the fact that in the $\Z_2\times \Z_2$ moduli-deformed fermionic constructions there are always $\N=2$ supersymmetric sectors, the question of how to avoid the decompactification problem arises. 

A solution for $\N=2$ supersymmetric models was proposed in Refs \cite{solving,solving2}. Actually, one may think that \\
($i$) \,if $\N=2$ is realized as a spontaneous breaking of $\N=4$ and\\
($ii$) if $\N=4$ is recovered when the large volume under consideration is sent to infinity,\\
then, instead of diverging as the volume, the threshold corrections should vanish in this limit. This expectation happens to be almost true, in the sense that in Eq. (\ref{D}), one finds $C^i=0$ and we are left with the mild, not dangerous, logarithmic divergence. In terms of orbifolds, the $\N=4\to \N=2$ spontaneous susy breaking is implemented by a $\Z_2$ action that is  free. In Sect. \ref{d-pb}, we will review the extension of this mechanism to the case of $\Z_2\times \Z_2$ moduli-deformed fermionic constructions, where $\N=1$ is further spontaneously broken to $\N=0$ by a SSS mechanism \cite{N=0thresh}.  However, due to the free action of (at least) one of the two $\Z_2$ actions, the resulting models are incompatible with the requirement of chirality of the massless spectrum, already at the level of $\N=1$. 


\subsection{Effective potential}

The 1-loop effective potential induced by the total spontaneous breaking of $\N\ge 1$ supersymmetries by a SSS mechanism is proportional to $m_{3/2}^4$, up to corrections of order $e^{-cM_{\rm s}/m_{3/2}}$, where $c=\O(1)$. This can be seen in heterotic and type II string orbifold models or moduli-deformed fermionic constructions. The $\N\ge 1\to \N=0$ spontaneous breaking can be implemented  \via  a $\Z_2$ shift along some large internal directions (this shift should be distinguished from some eventual $\Z_2\times \Z_2$ orbifold action). For simplicity, we again suppose that a single direction of radius $R$ is large and involved in the susy breaking. 

In this case, the winding modes along this direction as well as the ``twisted'' states (\ie strings closed up to the shift) have squared masses of order $R^2$, which implies their contributions to the vacuum amplitude are exponentially suppressed. We are left with the evaluation of a trace over the KK states of the ``untwisted'' sector, the original spectrum of the parent model. $(1+(-1)^G)/2$ is inserted in the trace, where $G$ is the  shift generator. The parent model being supersymmetric, the non-trivial result arises from the $(-1)^G\equiv(-1)^m$ insertion that leads to the KK contribution 
\begin{equation}
\label{kk}
\sum_m(-1)^m e^{-\pi \tau_2({m\over R})^2}={R\over \sqrt{\tau_2}}\sum_{\tilde m}e^{-\pi {R^2\over \tau_2}(\tilde m+{1\over 2})^2},
\end{equation}
where $\tau\equiv \tau_1+i\tau_2$ is the Teichmuller parameter of the genus-1 surface and the r.h.s. is obtained by Poisson resummation. This expression shows the very important fact that in the integral over the fundamental domain $\F$, the region $\tau_2\lesssim R^2$ is exponentially suppressed. Some remarks follow~:
\begin{description}
\item[\footnotesize $\bullet$] The supermassive modes (they contribute as $e^{-\pi\tau_2 (M/M_{\rm s})^2}$, with $M\gtrsim M_{\rm s}$) as well as the non-level matched ones (they contribute for $\tau_2\simeq 1$) yield exponentially suppressed corrections. As claimed before about the threshold corrections, the dominant contributions arise from the pure KK states associated to the large internal space. 
\item[\footnotesize $\bullet$] Up to exponentially suppressed terms, the $\tau_2$-integral can be extended to the ray $\tau_2>0$. Note that this remains true in a pure KK field theory : The tower of states regularizes the UV and there is no need to introduce a UV cutoff for the vacuum energy to be finite, provided that the initial spectrum is supersymmetric.\footnote{This finiteness in the UV is similar to that encountered in the computation of the free energy of a system at finite temperature, provided that the zero temperature vacuum energy is finite, which is the case in field theory if the spectrum is supersymmetric.}
\item[\footnotesize $\bullet$] Finally, the  change of variable $\tau_2\to \tau_2/R^2$ in the  integral that  arises  in the 1-loop effective potential leads the announced result,   
\begin{eqnarray}
\label{M4}
\V_{\mbox{\scriptsize 1-loop}}\!&\displaystyle =(n_{\rm F}-n_{\rm B}){(e^\phi M_{\rm s})^4 \over (2\pi)^3}\sum_{\tilde m}\int_0^{+\infty}{d\tau_2\over 2\tau_2^3} {R\over \sqrt{\tau_2}}e^{-\pi {R^2\over \tau_2}(\tilde m+{1\over 2})^2}\nonumber \\
&  \!\!\! \!\!\! \!\! \!\!\!\!\!\!\!\! \!\!\! \!\!\!\displaystyle = \xi(n_{\rm F}-n_{\rm B})\left({e^\phi M_{\rm s}\over R}\right)^4= \xi(n_{\rm F}-n_{\rm B})\, m_{3/2}^4\, ,
\end{eqnarray}
\end{description}
where $n_{\rm F}$ and $n_{\rm B}$ count the numbers of massless fermions and bosons in the parent theory that remain massless after breaking, while $\xi$ is a positive constant. More generally, when $n$ large internal directions are involved in the susy breaking, $\xi$ is a positive function of the complex structure moduli of the associated $n$-dimensional space.   

Having $m_{3/2}$ small compared to the string scale guarantees that $\V_{\mbox{\scriptsize 1-loop}}$ yields corrections to the observable sector tree level masses to be small. Moreover, if $n_{\rm F}>n_{\rm B}$, the run away behavior of $m_{3/2}$ induces naturally the desired ``gauge hierarchy'', $m_{3/2}\ll M_{\rm s}$. This happens dynamically as the Universe expands \cite{CosmologicalTerm}, during the cosmological era that precedes the electroweak phase transition at which the stabilization of $m_{3/2}$ is expected to occur, due to radiative corrections  \cite{radiative}. 

It is important to stress again that the  scaling  $m_{3/2}^4$ arises when the infinite towers of light KK states are taken into account. To appreciate this fact and its consequences for the mass corrections to be small, one can compare the situation with MSSM-like models.  The latter are seen as effective field theories, defined in rigid spacetime, with $\N=1$ supersymmetry softly broken. In this case, the tree level potential $\V_{\rm tree}$ contains mass terms at most of the order of the electroweak scale, to which one adds at 1-loop
\begin{equation}
\V_{\mbox{\scriptsize 1-loop}}=\Str \int^{\Lambda\over \sqrt{2}} {d^3 \mbox{\bf k}\over (2\pi)^3}{1\over 2}\sqrt{\mbox{\bf k}^2+M^2}\, .
\end{equation}
In this expression, $M$ is the tree level mass operator and $\Lambda/\sqrt{2}>\abs\abs\mbox{\bf k}\abs\abs$ is a UV cutoff of the order of the GUT, string or Planck scale. The integral over the 3-momentum  $\mbox{\bf k}$ can easily be computed explicitly. However, it is more illuminating to expend the result in powers of $M/\Lambda$, a fact that is allowed since the mass spectrum is bounded by $\Lambda$  :
\begin{equation}
\V_{\mbox{\scriptsize 1-loop}}={1\over 64\pi^2}\left( \Lambda^4\Str M^0+2\Lambda^2 \Str M^2+\Str M^4\ln\left({M^2\over 2\Lambda^2}\right)-{1\over2}\Str M^4+\O\left(\Str {M^6\over \Lambda^2}\right)\right).
\end{equation}
In this expression, the quartic divergence in $\Lambda$ is actually absent, due to equal numbers of bosons and fermions, $\Str M^0=0$. In rigid spacetime, the problem of having a very large cosmological constant is not addressed and $\Str M^2$ is allowed to be non-trivial. However, in softly broken supersymmetry, this term turns out  to be field-independent. Since $m_{3/2}$ in this context is a parameter and not a field,  this means  in practice that $\Str M^2$ is independent of the Higgs field. Thus, there is no loop correction of order $\Lambda^2$ to the Higgs squared mass and the latter gets only contributions of order $m_{3/2}^2$ from the remaining terms in $\V_{\mbox{\scriptsize 1-loop}}$.  For small enough susy breaking scale, the gauge hierarchy between the Higgs's VEV (and therefore the Standard Model's masses) and the cutoff is then guaranteed. 

However, at a more fondamental level, the derivation of MSSM-like models from $\N=1$ no-scale supergravities raises new constraints. When $\N=1$ (or $\N=2$) local supersymmetry is spontaneously broken to $\N=0$, helicity supertraces arguments yield \cite{strM2}
\begin{equation}
\Str M^0=0\, ,\qquad \Str M^2=c_2\, m_{3/2}^2 \, , \qquad \Str M^4=c_4\, m_{3/2}^4 \, ,
\end{equation}
for some scalar field-dependent $c_2$ and $c_4$ (independent of $m_{3/2}$). Note that $c_2$ takes into account the contributions arising from  the observable (MSSM-like) spectrum, the gravitational sector, as well as the hidden sector. When $c_2\neq 0$, the quadratically divergent $\Lambda^2$-term induces \\
($i$) a very large contribution to the cosmological constant we now have to face\\
($ii$) and a destabilization of the gauge hierarchy \cite{FKZ} : $m_{3/2}$ is now a field and minimizing $\V_{\mbox{\scriptsize 1-loop}}$ with respect to it yields a VEV $\langle m_{3/2}\rangle$ either equal to 0 (no supersymmetry breaking) or of order $\Lambda$ (no hierarchy). \\
Thus, phenomenologically viable supergravity theories in which loop corrections take into account a finite number of degrees of freedom require $c_2=0$, which is a strong constraint on the observable and non-observable spectra, referred as  ``Large Hierarchy Compatible'' (LHC) condition \cite{FKZ}.

To summarize, the no-scale models describe the total spontaneous breaking of $\N$ supersymmetries, with the lightest gravitino mass $m_{3/2}$ parametrizing at tree level a flat direction of a positive semi-definite potential. When the susy breaking arises from a (stringy) Scherk-Schwarz mechanism in string theory or in KK no-scale supergravity,  provided the gravitino mass is small compared to the string scale, there are infinite towers of light KK states that must be taken into account in the loop corrections. 
This yields a 1-loop effective potential that scales as $(n_{\rm F}-n_{\rm B})\, m_{3/2}^4$, up to corrections of order $e^{-cM_{\rm s}/m_{3/2}}$. If the gauge hierarchy is  stable, the cosmological constant is however still very large, compared to the currently observed one, since $(M_{\rm s}/R)^4=\O(10^{-56})M^4_{\rm P}\gg 10^{-120}M^4_{\rm P}$. We are thus invited to consider a subclass  of ``super no-scale models'' \cite{snc} characterized by $n_{\rm F}=n_{\rm B}$, which automatically go beyond the $m_{3/2}^4$ scaling. In this case, the 1-loop effective potential is essentially vanishing, being actually of  order $10^{-(c10^{13})}$. In other words, in super no-scale models, the no-scale structure persists at 1-loop, namely flatness of the direction parametrized by $m_{3/2}$ in a positive semi-definite effective potential, as long as $m_{3/2}<M_{\rm s}$. 

Since $n_{\rm F}$ and $n_{\rm B}$ count the total number of massless fermions and bosons, for a given observable sector, the relation $n_{\rm F}=n_{\rm B}$ imposes a constraint on the hidden sector, which seems in the spirit of the LHC condition. However, $n_{\rm F}-n_{\rm B}$ is not  strictly speaking a constant, since it depends on moduli fields.
In Sect. \ref{sns}, we show that among the models of Sect. \ref{d-pb}, where no decompactification problem arises,  the 1-loop effective potential in some cases drives dynamically the moduli that are {\em not involved} in the susy breaking to enhanced gauge symmetry points, such that the extended value of  $n_{\rm B}$ equals $n_{\rm F}$ \cite{snc}. These models are thus the first examples of super no-scale models. They are realized in moduli-deformed fermionic construction, where susy is spontaneously broken to $\N=0$ by a SSS mechanism. 

Note that pushing $m_{3/2}$ towards $M_{\rm s}$, the KK masses approach those of generic massive string states. The latter now contribute substantially to  $\V_{\mbox{\scriptsize 1-loop}}$, which does not {\em a prioti} vanish anymore. However, it is remarkable that in the examples of super no-scale models of Sect. \ref{sns}, no Hagedorn-like transition arises for any susy breaking scale. In other words, contrary to the case of temperature-like susy breaking, there are no state admitting winding numbers along directions involved in the susy breaking that become tachyonic  \cite{snc}. 


\section{A solution to the decompactification problem}
\label{d-pb}

In this section, we consider gauge  couplings in models, where a SSS mechanism implements an $\N=1\to \N=0$ spontaneous breaking at a low scale. Our aim is to find  conditions that avoid the decompactification problem \cite{N=0thresh}.


\subsection{The models}

In heterotic string in four dimensional Minkowski space, the gauge coupling threshold corrections  at 1-loop take the form \cite{thresholds,universality}
\begin{equation}
\Delta^i=\int_{\cal F}{d^2\tau \over \tau_2}\!
\left(
{1\over 2}\sum_{a,b}{\cal Q}[^a_b](2v)
\left({\cal P}_i^2(2\bar w) - {k^{i}\over 4\pi{\tau_2}}\right)\! \tau_2 \, {Z}[{^a_b}](2v,2\bar w)-b^i
\right)\!\!\Bigg\abs_{v=\bar w=0}
+b^i\log {2\,e^{1-\gamma}\over \pi\sqrt{27}} \, ,
\label{2}
\end{equation}
where ${Z}[^a_b](2v,2\bar w)$ is a refined partition function for given spin structures $a,b\in\Z_2$.  ${\cal P}_i(2\bar w)\equiv -\partial^2_{\bar w}/\pi^2\equiv i\partial_{\bar \tau}/\pi$ acts on the right-moving sector as the squared charge operator of the gauge group factor $G^i$, while ${\cal Q}[^a_b](2v)$ acts on the left-moving sector as the helicity operator,\footnote{Our conventions for the Jacobi functions $\theta[^a_b](\nu\abs\tau)$ (or $\theta_{\alpha}(\nu\abs\tau)$, $\alpha=1,\dots,4$) can be found in \cite{KiritsisBook}.} 
\begin{equation}
\label{Qop}
{\cal Q}[^a_b](2v) = {1\over 16\pi^2}\, {\partial_v^2(\theta[^a_b](2v))\over \theta[^a_b](2v)}-{i\over\pi}\, \partial_\tau\log\eta\equiv {i\over \pi}\, \partial_{\tau}\!\left(\log {\theta[^a_b](2v)\over\eta} \right) .
\end{equation}

From now on, we consider  $\Z_2\times \Z_2$ moduli-deformed fermionic models, where $\N=1$ is spontaneously broken by a SSS mechanism. The associated genus-1 refined partition function is
\begin{eqnarray}
\label{Z}
   Z(2v,2\bar w)=\!&\displaystyle {1\over \tau_2(\eta \bar\eta)^2}\,{1\over 2} \sum_{a,b}\,  {1\over 2}  \sum_{H_1,G_1} \, {1\over 2}  
\sum_{H_2,G_2} e^{i\pi (a+b+ab)}\, {\theta[^a_b ](2v) \over \eta}\, 
{\theta[^{a+H_1}_{b+G_1}] \over \eta}\,{\theta[^{a+H_2}_{b+G_2}] \over \eta}\, {\theta[^{a+H_3}_{b+G_3}] \over \eta}\nonumber\\
&\displaystyle\!\!\!\!\!\!\!\times \,  {1\over 2^N}\sum_{h_I^i,g_I^i}S_L\Big[{}^{a,\, h^i_I,\,  H_I}_{b,\,  g^i_I,\, G_I}  \Big]  \, Z_{2,2}\Big[{}^{h_1^i}_{g_1^i} 
 \Big \abs {}^{H_1}_{G_1} \Big]\, 
 Z_{2,2}\Big[{}^{h_2^i}_{g_2^i} \Big \abs{}^{H_2}_{G_2} \Big]\,
 Z_{2,2}\Big[{}^{h_3^i}_{g_3^i} \Big \abs{}^{H_3}_{G_3} \Big] \,
 Z_{0,16}\Big[{}^{h_I^i , \,  H_I}_{g_I^i ,\, G_I}  \Big] \;\!\!(2\bar w)\,  ,
\label{partition}
\end{eqnarray}
where our notations are as follows :
\begin{description}
\item[\footnotesize $\bullet$] The $Z_{2,2}$ conformal blocks are the contributions of the three internal 2-tori, whose coordinates are shifted and twisted. Denoting the $\Z_2$-shifts characters as $(h^i_I, g^i_I)$, $i=1,2$, $I=1,2,3$ and the $\Z_2$-twist characters as $(H_I,G_I)$, where $(H_3,G_3)\equiv (-H_1-H_2,-G_1-G_2)$, we have
$$
\!\!\!\!\!\!\!\!\!\!\!\!\!\!\!\!\!\!Z_{2,2}\Big[{}^{h_I^1,\, h_I^2} _{g_I^1,\, g_I^2} \Big \abs {}^{H}_{G} \Big]\!=\left\{
\begin{array}{ll}
\displaystyle {\Gamma_{2,2}\Big[{}^{h_I^1,\, h_I^2} _{g_I^1,\, g_I^2} \Big]\!(T_I,U_I)\over (\eta\bar\eta)^2}\, ,\phantom{\delta_{\huge\abs{}^{h_I^1\;\;H_I}_{g_I^1\;\; G_I}\big\abs,0}}&\mbox{when }  (H_I, G_I)=(0,0)\, ,\\ 
\displaystyle {4\eta \bar \eta\over\ \theta[^{1-H_I}_{1-G_I} ]\, \bar\theta[^{1-H_I}_{1-G_I}]}\; 
\delta_{\big\abs{}^{h_I^1\;\;H_I}_{g_I^1\;\; G_I}\big\abs,0 \,\mbox{\scriptsize mod} \,2}\;\delta_{\big\abs{}^{h_I^2\;\;H_I}_{g_I^2\;\; G_I}\big\abs,0 \,\mbox{\scriptsize mod} \,2}\, , &\mbox{when }(H_I, G_I)\neq(0,0)\, ,
\end{array}
\right.
$$
\vspace{-1.31cm}
\begin{equation}\label{toto}\phantom{\delta_{\huge\abs{}^{h_I^1\;\;H_I}_{g_I^1\;\; G_I}\big\abs,0}}\end{equation}
where $\Gamma_{2,2}$ is a shifted lattice that depends on the $T_I$ and $U_I$ \Ka and complex structure moduli of the $I^{\rm th}$ 2-torus.
\item[\footnotesize $\bullet$] Linear contraints on the shifts $(h^i_I,g^i_I)$ and twists $(H_I,G_I)$ may be imposed, leaving effectively $N$ independent shift characters. \item[\footnotesize $\bullet$] $Z_{0,16}$ denotes the contribution of the 32 extra right-moving worldsheet fermions. Its dependance on the shift and twist characters may generate discrete Wilson lines, which break partially $E_8\times E_8$ or $SO(32)$. 
\item[\footnotesize $\bullet$] The first line contains the contribution of the spacetime  light-cone bosons, with that of the left-moving  fermions.  
\item[\footnotesize $\bullet$] $S_L$ is a conformal block-dependent sign that implements the SSS mechanism. If $S_L$ is identically equal to 1, the model is  $\N=1$ supersymmetric. However, a non-trivial $S_L$ correlating the spin structure $(a,b)$ to some lattice charges $(h^i_I,g^i_I)$ implements the $\N=1\to\N=0$ spontaneous breaking.  
\end{description}

For the decompactification problem not to arise, we impose one of the three $\N=2$ sectors to be realized as a spontaneous breaking of $\N=4$. This can be implemented by demanding the $\Z_2$ action, whose characters are $(H_2,G_2)$, to be free. Its generator twists the $2^{\rm nd}$ and $3^{\rm rd}$ 2-tori (\ie the directions $X^6,X^7,X^8,X^9$ in bosonic language) and shifts some directions of the $1^{\rm st}$ 2-torus, say $X^5$. To simplify our discussion, we take the generator of the second $\Z_2$, whose characters are $(H_1,G_1)$, to not be free : It twists the $1^{\rm st}$ and $3^{\rm rd}$ 2-tori, and fixes the $2^{\rm nd}$ one. Similarly, we suppose the product of the two generators, whose characters are $(H_3,G_3)$, twists the $1^{\rm st}$ and $2^{\rm nd}$ 2-tori, and fixes the $3^{\rm rd}$ one. These restrictions impose the moduli $T_2,U_2$ and $T_3,U_3$ not to be far from $1$, in order to avoid  the decompactification problem  to occur from the other two $\N=2$ sectors. Thus, the final SSS $\N=1\to \N=0$ spontaneous breaking must involve the moduli $T_1,U_1$ only, for the gravitino mass to be light. This breaking is implemented \via a shift along the $1^{\rm st}$ 2-torus, say $X^4$, and a non-trivial choice of $S_L$. Therefore, the sector $(H_1,G_1)=(0,0)$  realizes the pattern of spontaneous breaking $\N=4\to \N=2\to \N=0$, while the two other $\N=2$ sectors, which are associated to the $2^{\rm nd}$ and $3^{\rm rd}$ 2-tori, are independent of $T_1$ and $U_1$ and thus  remain supersymmetric. As a result, we have in the two following independent modular orbits :
\begin{eqnarray}
\label{S}
&S_L=e^{i\pi(ag^1_1+bh^1_1+h^1_1g^1_1)}\,,&\when\quad (H_1,G_1)=(0,0)\, ,\nonumber \\
&\!\!\!\!\!\!\!\!\!\!\!\!\!\!\!\!\!\!\!\!\!\!\!\!\!\!\!\!\!\!\!\!\!\!\!\!\!\!\!\!\!\!\!\!\!\!\!\!\!\!\!\!S_L=1\, ,&\when \quad(H_1,G_1)\neq(0,0)\, .
\end{eqnarray}

Following the conventions chosen in the above paragraph, we have $(h^2_1,g^2_1)\equiv (H_2,G_2)$ and the $1^{\rm st}$ 2-torus lattice takes the explicit form\footnote{There are actually 6 distinct choices of shifts along the $1^{\rm st}$ 2-torus to implement the $\N=4\to \N=2$ and $\N=1\to \N=0$ spontaneous breaking. However, they are all equivalent, up to permutations of the sectors to be discussed in the next section \cite{N=0thresh}.}  
\begin{eqnarray}
&\displaystyle\Gamma_{2,2}\Big[{}^{h_1^1,\, H_2} _{g^1_1,\, G_2} \Big](T_1,U_1) =\sum_{m^i,n^i}(-)^{m^1g^1_1+m^2G_2}\, &\displaystyle e^{2i\pi\tau\left[m^1\left(n^1+{1\over 2}{h^1_1}\right)+m^2\left(n^2+{1\over 2}{H_2}\right)\right]}\,\times\nonumber\\
&&e^{-{\pi\tau_2\over \Im T_1\Im  U_1}\left\abs T_1\left(n^1+{1\over 2}{h^1_1}\right)+T_1U_1\left(n^2+{1\over 2}{H_2}\right)-U_1m^1+m^2\right\abs^2}.
\end{eqnarray}
This expression can be used to find the squared scales of spontaneous $\N=4\to \N=2$ and $\N=2\to \N=0$ breaking,\footnote{We display them for $\Re(U_1)\in (-1,1]$.}  
\begin{equation}
\label{m32}
{M^2_{\rm s}\over \Im T_1\, \Im U_1}\; , \qquad m^2_{3/2}={\abs U_1\abs M^2_{\rm s}\over \Im T_1\, \Im U_1}\, ,
\end{equation}
the latter being identified with the gravitino mass squared in the full $\N=0$ theory. To impose these scales to be small compared to the string scale, we consider the regime $\Im T_1\gg 1$ and $U_1=\O(1)$. 


\subsection{Conformal block by conformal block analysis}
\label{sbys}

The threshold corrections can be analyzed in each conformal block \cite{N=0thresh}. Starting with those where $(H_1,G_1)=(0,0)$, the discussion is facilitated by summing over the spin structures. Focussing on the relevant parts of the refined partition fonction $Z$, we have
\begin{eqnarray}
&\displaystyle {1\over 2}\sum_{a,b}e^{i\pi (a+b+ab)}& \!\!e^{i\pi(ag^1_1+bh^1_1+h^1_1g^1_1)}\,\theta[^a_b ] (2v)\, \theta [^a_b ]\, \theta[^{a+H_2}_{b+G_2} ]\, \theta[^{a-H_2}_{b-G_2} ] \nonumber \\
&&\qquad\qquad =e^{i\pi \left(h^1_1g^1_1+G_2(1+h^1_1+H_2)\right)}\, \theta[^{1-h^1_1}_{1-g^1_1} ]^2(v) \,  \theta[^{1-h^1_1+H_2}_{1-g^1_1+G_2} ]^2(v)  \,  ,
\label{theta}
\end{eqnarray}
which shows how many odd $\theta_1(v)\equiv \theta[^1_1](v)$ functions (or equivalently how many fermionic zero modes in the path integral) arise for given characters $(h^1_1,g^1_1)$ and $(H_2,G_2)$. \\

\vspace{1.4cm}
\noindent {\large \em Conformal block $A$ : $(h^1_1,g^1_1)=(0,0)$, $(H_2,G_2)=(0,0)$} 

This block  is  proportional to $\theta[^1_1]^4(v)=\O(v^4)$ and  arises, up to an overall $1/2^3$ normalization factor, from the $\N=4$ spectrum of the parent supersymmetric theory. The action of the helicity operator being $\O(v^2)$ (see Eq. (\ref{Qop})), the contribution to the thresholds is trivial, $\Delta^i_A=0$.\\

\noindent {\large \em Conformal blocks $B$ : $(h^1_1,g^1_1)\neq(0,0)$, $(H_2,G_2)=(0,0)$} 

They are proportional to $\theta [^{1-h^1_1}_{1-g^1_1}]^4(v)=\O(1)$ and the action of the helicity operator yields a non-trivial contribution $\Delta_B^i$ to the gauge coupling thresholds. However, the winding number along the very large periodic direction $X^4$ being $2n^1+h^1_1$, the blocks with $h^1_1=1$ are exponentially suppressed, compared to the remaining one, with $(h^1_1,g^1_1)=(0,1)$. Up to an overall factor $1/2^2$, the latter is the contribution of the spectrum considered in the conformal block $A$, but in the spontaneously broken phase, $\N=4\to \N=0$. It leads to 
\begin{equation}
\Delta^i_B={b^i_B\over 4}\Delta_B-{k^i\over 4}Y_B\, ,
\end{equation}
where $\Delta_B$ and $Y_B$ are dominated by the light KK states associated to the $1^{\rm st}$ 2-torus,  
\begin{eqnarray}
&\Delta_B\!&=-\ln\left({\pi^2\over 4}
|\theta_2(U_1)|^4\,\Im T_1\,\Im U_1  \right)+{\cal O}\left(e^{-c\,\sqrt{\Im T_1}}\right) ,\nonumber\\
& Y_B\!&=-{2+d_{G_B}-n_{{\rm F}_B}\over 3\pi^3}\, {1\over  \Im T_1}\, E_{(1,0)}(U_1\abs\, 2)+{\cal O}\left(e^{-c\,\sqrt{\Im T}}\right).
\end{eqnarray}
In $Y_B$, we use ``shifted real analytic  Eisenstein series'', which are defined as
\begin{equation}
E_{(g_1,g_2)}(U\abs \, s)={\sum_{\tilde m_1,\tilde m_2}}^{\!\!\!\!\prime}\,{(\Im U)^s\over \abs\tilde m_1+{1\over 2}g_1+(\tilde m_2+{1\over 2}g_2)U\abs^{2s}}\, .
\end{equation}
As expected, for large $1^{\rm st}$ 2-torus volume $\Im T_1$, the threshold correction $\Delta_B^i$ scales like $\ln\Im T_1$ and not the volume itself.
Moreover, it contains a gauge group factor-dependent contribution proportional to a $\beta$-function coefficient, which involves Casimir coefficients,
\begin{equation}
b^i_B=-{8\over3}\left\{C({\cal A}_B^i)-C({\cal R}_B^i) \right\}.
\end{equation}
The term $-{8\over 3}C({\cal A}_B^i)$ comes from the bosons of some initially massless $\N=4$ vector multiplets in the parent $\N=4$ model that remain massless. These bosons (1 vector and 2 real scalars) are in the adjoint representation ${\cal A}^i_B$ of a gauge group factor $G^i_B$. The contribution ${8\over 3}C({\cal R}^i_B)$ arises from the fermions  of some initially massless $\N=4$ vector multiplets in the parent theory that remain massless. They are 4 Majorana  fermions in a spinorial representation ${\cal R}_B^i$ of $G_B^i$. On the contrary, $Y_B$ is a gauge group factor-independent contribution. It is proportional to the difference between the numbers of massless bosons and massless fermions, $8(2+d_{G_B}-n_{{\rm F}_B})$. In this expression, the 2 is associated to the  spacetime light-cone coordinates, $d_{G_B}$ is the dimension of the full gauge group realized in the conformal block $B$ and $8n_{{\rm F}_B}$ is the number of massless fermionic degrees of freedom in this block.
\\

\vspace{0.3cm}
\noindent {\large \em Conformal blocks $C$ : $(h^1_1,g^1_1)=(0,0)$, $(H_2,G_2)\neq(0,0)$} 

 They are proportional to $\theta[^1_1](v)^2\theta[^{1-H_2}_{1-G_2}]^2(v)=\O(v^2)$ and the helicity operator yields a contribution $\Delta_C^i$ to the thresholds. Since the winding numbers along the large periodic direction $X^5$ are $2n^2+H_2$, the blocks with $H_2=1$ are exponentially suppressed, compared to the remaining one, with $(H_2,G_2)=(0,1)$. Up to an overall factor $1/2^2$, this block is the contribution of a spectrum realizing the spontaneous $\N=4\to\N_C=2$ breaking, which yields
\begin{equation}
\Delta^i_C={b^i_C\over 4}\Delta_C-{k^i\over 4}Y_C\, ,
\end{equation}
where the light KK states dominate :
\begin{eqnarray}
&\Delta_C\!&=  -\ln\left({\pi^2\over 4}
|\theta_4(U_1)|^4\,\Im T_1\,\Im U_1  \right)+{\cal O}\left(e^{-c\,\sqrt{\Im T_1}}\right) ,\nonumber\\
& Y_C\!&=-{2+n_{{\rm V}_C}-n_{{\rm H}_C}\over 3\pi^3}\, {1\over  \Im T_1}\, E_{(0,1)}(U_1\abs\, 2)+{\cal O}\left(e^{-c\,\sqrt{\Im T}}\right).
\end{eqnarray}
As expected, the large $\Im T_1$ behavior is logarithmic. The gauge group factor-dependent part depends on the $\beta$-function coefficient  
\begin{equation}
b^i_C=-2\left\{C({\cal A}_C^i)-C({\cal R}_C^i) \right\},
\end{equation}
which is obtained from the massless $\N_C=2$ vector multiplets feeling the adjoint representation ${\cal A}_C^i$ of a gauge group factor $G_C^i$, and from charged hypermultiplets in the representation ${\cal R}_C^i$ of $G_C^i$. In the gauge group factor-independent contribution $Y_C$,  $n_{{\rm V}_C}$ and $n_{{\rm H}_C}$ are the total numbers of massless vector multiplets and hypermultiplets  realized in the block $C$.
\\

\noindent {\large \em Conformal blocks $D$ : $(h^1_1,g^1_1)=(H_2,G_2)\neq(0,0)$} 

 They are proportional to $\theta[^{1-H_2}_{1-G_2}]^2(v)\theta[^1_1](v)^2=\O(v^2)$ and contribute to the thresholds in a way similar to that of the conformal blocks $C$. The dominant contribution arises again for $(H_2,G_2)=(0,1)$ and is associated, up to an overall factor $1/2^2$, to a spectrum realizing a spontaneous $\N=4\to \N_D=2$  breaking. The latter is realized \via a $\Z_2$ action, whose generator twists the $2^{\rm nd}$ and $3^{\rm rd}$ 2-tori, and effectively shifts both directions $X^4$ and $X^5$. Note that the $\N_C=2$ and $\N_D=2$ supersymmetries are distinct. The contribution to the thresholds is 
\begin{equation}
\Delta^i_D={b^i_D\over 4}\Delta_D-{k^i\over 4}Y_D\, ,
\end{equation}
where
\begin{eqnarray}
&\Delta_D\!&=-\ln\left({\pi^2\over 4}
|\theta_3(U_1)|^4\,\Im T_1\,\Im U_1  \right)+{\cal O}\left(e^{-c\,\sqrt{\Im T_1}}\right) ,\nonumber\\
& Y_D\!&=-{2+n_{{\rm V}_D}-n_{{\rm H}_D}\over 3\pi^3}\, {1\over  \Im T_1}\, E_{(1,1)}(U_1\abs\, 2)+{\cal O}\left(e^{-c\,\sqrt{\Im T}}\right),
\end{eqnarray}
with logarithmic large $\Im T_1$ behavior. 
The $\beta$-function coefficient  
\begin{equation}
b^i_D=-2\left\{C({\cal A}_D^i)-C({\cal R}_D^i) \right\}
\end{equation}
arises from the massless $\N_D=2$ vector multiplets in the adjoint representation ${\cal A}_D^i$ of a gauge group factor $G_D^i$, and from  charged hypermultiplets in the representation ${\cal R}_D^i$ of $G_D^i$. $n_{{\rm V}_D}$ and $n_{{\rm H}_D}$ are the numbers of massless vector multiplets and hypermultiplets  realized in the block $D$.\\

\noindent {\large \em Conformal blocks $E$ : $h^1_1G_2-g^1_1H_2\neq 0$} 

 Since the defining conditions of the conformal blocks $A, \dots, D$ are the solutions to the equation $\Big\abs{}^{h^1_1\;\,H_2}_{g^1_1\; \,G_2}\Big\abs=0$, the remaining blocks $E$ have non-vanishing determinants. Their contributions are proportional to $\theta[^{1-h^1_1}_{1-g^1_1} ]^2(v) \theta[^{1-h^1_1+H_2}_{1-g^1_1+G_2} ]^2(v)=\O(1)$ and the action of the helicity operator gives a non-trivial correction $\Delta^i_E$ to the thresholds. However, non-vanishing of the determinant implies $(h^1_1,H_2)\neq (0,0)$, which shows that all states contributing in these blocks have non-trivial winding numbers along $X^4$, $X^5$ or both. Therefore, their contributions are exponentially suppressed, $\Delta^i_E={\cal O}\left(e^{-c\,\sqrt{\Im T_1}}\right)$.\\

After having analyzed all conformal blocks satisfying $(H_1,G_1)=(0,0)$, we  proceed with the study of the modular orbit $(H_1,G_1)\neq(0,0)$. The SSS sign $S_L$ is now trivial, as indicated in Eq. (\ref{S}). Since the $1^{\rm st}$ 2-torus is twisted, these blocks are independent of the moduli $T_1,U_1$ and thus the susy breaking scale. They can be analyzed as in the case of $\Z_2\times\Z_2$ $\N=1$ supersymmetric fermionic models. Actually, summing over the spin structures, the relevant terms in the refined partition function $Z$ become 
\begin{eqnarray}
&\displaystyle {1\over 2}
\sum_{a,b}e^{i\pi (a+b+ab)}\!\!& \!\theta[^a_b ] (2v)\, \theta [^{a+H_1}_{b+G_1}]\, \theta[^{a+H_2}_{b+G_2}]\, \theta[^{a-H_1-H_2}_{b-G_1-G_2} ]=\nonumber \\
&&e^{i\pi (G_1+G_2)(1+H_1+H_2)}\, \theta[^{1}_{1}](v) \, 
 \theta[^{1-H_1}_{1-G_1} ](v)\, \theta[^{1-H_2}_{1-G_2}](v) \, 
 \theta[^{1+H_1+H_2}_{1+G_1+G_2} ](v) \, ,
\end{eqnarray}
which invites us to split the discussion in three parts. 
\\

\noindent {\large \em Conformal blocks of the 2$^{nd}$ plane : $(H_2,G_2)=(0,0)$} 

 The reference to the $2^{\rm nd}$ plane means that the  $2^{\rm nd}$ internal 2-torus is fixed by the non-free action of the $\Z_2$, whose characters are $(H_1,G_1)$. These blocks are proportional to $\theta[^1_1]^2(v)\theta[^{1-H_1}_{1-G_1}]^2(v)=\O(v^2)$ and the helicity operator yields a non-trivial correction $\Delta_2^i$ to the thresholds. Adding the trivial contribution arising from the conformal block $A$, up to an overall factor $1/2$, one has to compute the threshold corrections arising from an $\N_2=2$ supersymmetric spectrum, with $\N_2=2$ $\beta$-function coefficient $b^i_2$.  In the case of $(2,2)$ superconformal symmetry on the worldsheet, the result is  
\begin{equation}
\Delta^i_2={b^i_2\over 2}\Delta(T_2,U_2)-{k^i\over 2} Y(T_2,U_2)\, ,
\label{D2}
\end{equation}
where 
\begin{eqnarray}
&\Delta(T,U)\!&=-\ln\left(4\pi^2 \big\abs\eta(T)\big\abs^4 \,\big\abs\eta(U)\big\abs^4\, \Im T\,  \Im U\right),\phantom{\delta_{\large \abs{}^{h_I^1\;\;H_I}_{g_I^1\;\; G_I}\big\abs,0}}\nonumber\\
& Y(T,U)\!&={1 \over 12}\int_{\cal F}{d^2\tau\over \tau_2}\,
\Gamma_{2,2}(T,U) \left[ \Big(\bar E_2-{3\over \pi\tau_2}\Big){\bar E_4 \bar E_6\over \bar\eta^{24}}-\bar \jmath+1008 \right].
\end{eqnarray}
In these expressions, $\Gamma_{2,2}$ is the unshifted lattice associated to the $2^{\rm nd}$ 2-torus, $E_{2,4,6}=1+\O(q)$ are holomorphic Eisenstein series of modular weights 2,4,6, where $q\equiv e^{2i\pi\tau}$, and $j={1\over q}+744+\O(q)$ is holomorphic and modular invariant. 
\\

\noindent {\large \em Conformal blocks of the 3$^{rd}$ plane : $(H_1,G_1)=(H_2,G_2)$} 

In this case $(H_3,G_3)=(0,0)$, which means that the $3^{\rm rd}$ 2-torus is fixed by the combined action of the generators of the two $\Z_2$'s. As in the case of the $2^{\rm nd}$ plane, these blocks are proportional to  $\theta[^1_1]^2(v)\theta[^{1-H_1}_{1-G_1}]^2(v)=\O(v^2)$ and the contribution $\Delta^i_3$ to the thresholds is of identical form, 
\begin{equation}
\Delta^i_3={b^i_3\over 2}\Delta(T_3,U_3)-{k^i\over 2} Y(T_3,U_3)\, ,
\label{D3}
\end{equation}
where $b^i_3$ is the associated $\N_3=2$ $\beta$-function coefficient. Notice that the  $\N_C=2$, $\N_D=2$, $\N_2=2$ and $\N_3=2$ supersymmetries we have encountered are all distinct. 
\\

\noindent {\large \em Conformal blocks with $\N=1$ : $H_1G_2-G_1H_2\neq 0$} 

 The remaining blocks have non-trivial determinant, $\big\abs{}^{H_1\;H_2}_{G_1\; G_2}\big\abs\neq 0$. Acting on them with the helicity operator, the result is proportional to 
\begin{equation}
\partial_v^2\Big(\theta[^1_1](v)\, \theta[^{1-H_1}_{1-G_1}](v)\, \theta[^{1-H_2}_{1-G_2}](v)\,  \theta[^{1+H_1+H_2}_{1+G_1+G_2}](v)\Big)\!\Big\abs_{v=0}\propto \partial_v^2\Big(\theta_1(v)\, \theta_2(v)\, \theta_3(v)\,  \theta_4(v)\Big)\!\Big\abs_{v=0}=0\, ,
\end{equation}
thanks to the oddness of $\theta_1(v)$ and evenness of $\theta_{2,3,4}(v)$.
Thus, these conformal blocks do not  contribute to the thresholds, $\Delta_{\N=1}^i=0$.


\subsection{Universal running gauge couplings}
\label{g1}

All the above results can be summarized by introducing moduli-dependent squared masses~\cite{N=0thresh},
\begin{eqnarray}
&&M^2_B={M^2_{\rm s}\over |\theta_2(U_1)|^4\, \Im T_1\, \Im U_1}\nonumber \\
&&M^2_C={M^2_{\rm s}\over |\theta_4(U_1)|^4\, \Im T_1\, \Im U_1}\nonumber \\
&&M^2_D={M^2_{\rm s}\over |\theta_3(U_1)|^4\, \Im T_1\, \Im U_1}\nonumber \\
&&M^2_I={M^2_{\rm s} \over 16\big\abs\eta(T_I)\abs^4 \, \big\abs\eta(U_I)\abs^4\, \Im T_I \, \Im U_I}=\O(M^2_{\rm s}), \quad I=2,3,
\end{eqnarray}
and absorbing the gauge group-independent contributions $Y(T_I,U_I)$ in a ``renormalized string coupling'' as \cite{universality} 
\begin{equation}
{16\, \pi^2\over g_{\rm renor}^2}={16\, \pi^2\over g_{\rm s}^2}-{1\over 2}Y(T_2,U_2)-{1\over 2}Y(T_3,U_3)\, .
\end{equation}
In terms of these notations, the running  effective gauge coupling at energy scale $Q^2 =\mu^2 {\pi^2\over 4}$ takes a universal form,
\begin{eqnarray}
\label{thfinal}
{16\, \pi^2\over g_{i}^2(Q)} = k^{i}{16\, \pi^2\over g_{\rm renor}^2} 
&\displaystyle -{b^i_{B}\over 4}\ln\left({Q^2\over Q^2+M^2_B}\right)
-{b^i_C\over 4}\ln\left({Q^2\over Q^2+M^2_C}\right)
-{b^i_{D}\over 4}\ln\left( {Q^2\over Q^2+M^2_D}\right)\phantom{\delta_{\huge\abs{}^{h_I^1\;\;H_I}_{g_I^1\;\; G_I}\big\abs,0}}\nonumber \\
 & \displaystyle -{b^i_{2}\over 2}  \ln\left({Q^2\over M^2_{2}} \right) 
 -{b^i_{3}\over 2}\ln\left( {Q^2\over M^2_{3}}\right)+\O\left({m^2_{3/2}\over M^2_{\rm s}}\right),
\end{eqnarray}
which depends only on five  model-dependent $\beta$-function coefficients. 
In this expression, we have shifted $M_{B,C,D}^2\to Q^2+M_{B,C,D}^2$ to implement the  thresholds at which the sectors $B$, $C$ or $D$ decouple, \ie when $Q$ exceeds $M_B$, $M_C$ or $M_D$. Thus, the expression for $g_i^2(Q)$ is valid up to $cM_{\rm s}$, at which massive states we have neglected start  to contribute substantially.

Before concluding this section, we would like to justify that at the $\N=1$ level, the $\Z_2\times \Z_2$ models, with at least one freely acting $\Z_2$, describe a  non-chiral spectrum. In general, in the $\Z_2\times \Z_2$ models, the chiral families arise from the fixed points. If the $\Z_2$, with characters $(H_2,G_2)$, is freely acting, then the $1^{\rm st}$ plane is not fixed (all states in this sector are massive) and chiral families may only arise from the $2^{\rm nd}$ and $3^{\rm rd}$ planes. Reversing the order of the two $\Z_2$ actions, the model can be seen as follows. The $\Z_2$, with characters $(H_1,G_1)$, realizes the breaking (``hard'' or spontaneous) from $\N=4$ to $\N=2$, while the $\Z_2$, with characters $(H_2,G_2)$, realizes the spontaneous breaking $\N=2\to \N=1$. However, the massless twisted spectrum associated to the $2^{\rm nd}$ and $3^{\rm rd}$ planes is independent of the modulus that is the order parameter of the spontaneous $\N=2\to \N=1$ breaking. Taking the (large volume) limit in which $\N=2$ is restored, one concludes that this spectrum respects exact $\N=2$ supersymmetry at tree level, and is therefore non-chiral. 

Moreover, implementing a final $\N=1\to \N=0$ spontaneous breaking \via a shift along the $1^{\rm st}$ 2-torus only (if the other two planes are fixed, this is required to avoid the decompactification problem), the degeneracy of the chiral superfields realizing the twisted hypermultiplets of the $2^{\rm nd}$ and $3^{\rm rd}$ planes cannot be lifted. 
 

\section{Super no-scale models}
\label{sns}

In this section, we still consider  moduli-deformed fermionic constructions, where $\N=4$, 2 or 1 supersymmetry is  spontaneously broken to $\N=0$ \via a SSS mechanism, and where the decompactification problem does not arise. Our aim is to show that in this setup, super no-scale models exist \cite{snc}. In other words, the no-scale structure persists  in some cases at 1-loop, for low supersymmetry breaking scale. We also show that these models do not develop tachyonic  instabilities at tree level, even when the supersymmetry breaking scale is high.


\subsection{Relevance of the conformal blocks $B$}

In four dimensional Minkowski space, our interest is focussed on the 1-loop effective potential, which takes in general the following form, in terms of the partition function $Z$,
\begin{equation}
\V_{\mbox{\scriptsize 1-loop}}= -{1\over (2\pi)^4}\int_\F {d^2\tau\over 2\tau_2^2}\, Z\abs_{v=\bar w=0}\, .
\end{equation}
In the $\Z_2\times \Z_2$ models considered in Sect. \ref{d-pb}, the generator that twists the $2^{\rm nd}$ and $3^{\rm rd}$ 2-tori is also shifting the direction $X^5$, while the final SSS $\N=1\to \N=0$ breaking is introduced as a shift along $X^4$. The $1^{\rm st}$ 2-torus moduli satisfy $\Im T_1\gg 1$, $U_1=\O(1)$, which guaranties $m_{3/2}\ll M_{\rm s}$, while the $2^{\rm nd}$ and $3^{\rm rd}$ 2-tori, which are fixed by the two remaining non-trivial group elements of $\Z_2\times \Z_2$, have moduli $T_2,U_2$ and $T_3,U_3$ not far from 1, for the decompactification problem not to arise. 

In these models, we have seen that the conformal blocks $A$, $C$, $D$ and those associated to the $2^{\rm nd}$ plane, $3^{\rm rd}$ plane, as well as the $\N=1$ blocks are supersymmetric. Thus, they do not contribute to the vacuum energy, $\V_{\mbox{\scriptsize 1-loop}}$. Moreover, among the remaining blocks $B$ and $E$, which are not supersymmetric, the second arise from states, with non-trivial winding numbers along $X^4$, $X^5$ or both. Therefore, the contribution of the blocks $E$  is exponentially suppressed, $\O(e^{-c\,\sqrt{\Im T_1}})$, compared to that of the blocks $B$, which is the only one  we need to evaluate in practice. The result~is 
\begin{equation}
\label{v1}
\V_{\mbox{\scriptsize 1-loop}}= {1\over 4}\, {8(n_{{\rm F}_B}-2-d_{G_B})\over 16\pi^7}\, {1\over (\Im T_1)^2}\, E_{(1,0)}(U_1\abs \,3)+{\cal O}\!\left(e^{-c\,\sqrt{\Im T_1}}\right),
\end{equation}
which is proportional to $m_{3/ 2}^4$, as explained in the paragraph before Eq. (\ref{M4}). The coefficient $8(n_{{\rm F}_B}-2-d_{G_B})$ is the number of massless fermions minus the number of massless bosons in the parent $\N=4\to \N=0$ theory. The overall factor $1/4$ is the  normalization to be added in the $\Z_2\times \Z_2$, $\N=1\to \N=0$ case. This normalization factor becomes $1/2$, in the $\N=2\to \N=0$ models realized with the single freely acting $\Z_2$, whose characters are $(H_2,G_2)$. In other words, in the present context, to show that some $\N=1$ or  $\N=2$ no-scale model is actually super no-scale, it is  enough to consider its parent $\N=4\to \N=0$ theory and show it satisfies $n_{\rm F}-n_{\rm B}\equiv8(n_{{\rm F}_B}-2-d_{G_B})=0$. 


\subsection{Examples of super no-scale models}

Motivated by the previous discussion, we are going to construct an $\N=4$ super no-sacle model. Our starting point is the  genus-1 partition function of the exact $\N=4$, $E_8\times E_8'$ heterotic string compactified on $T^2\times T^2\times T^2$, which takes the factorized form
\begin{equation}
\label{N4}
Z_{\N=4}= O^{(0)}_{2,2}\,  O^{(1)}_{2,2}\, O^{(2)}_{2,2}\, O^{(3)}_{2,2}\; {1\over 2}\sum_{a,b}(-)^{a+b+ab}\, {\theta[^a_b]^4\over \eta^4}\; {1\over 2}\sum_{\gamma,\delta}{\bar \theta[^\gamma_\delta]^8\over \bar \eta^8}\; {1\over 2}\sum_{\gamma',\delta'}{\bar \theta[^{\, \gamma'}_{\delta'}]^8\over \bar \eta^8}\, , 
\end{equation}
where $O^{(0)}_{2,2}$ denotes  the contribution of the spacetime light-cone  coordinates and the $O^{(I)}_{2,2}$, $I=1,2,3$, stand for the contributions of the $T^2$'s directions, 
\begin{equation}
O^{(0)}_{2,2} ={1\over \tau_2\, (\eta \bar \eta)^2}\,,\qquad  O^{(I)}_{22}={\Gamma_{2,2}(T_I,U_I)\over(\eta\bar \eta)^2} \, , \;\;I=1,2,3\,.
\end{equation}
Defining 
 the holomorphic $SO(2N)$ characters as follows, 
\begin{eqnarray}
&&\displaystyle O_{2N}={\theta[^0_0]^N+\theta[^0_1]^N\over 2\eta^N}\, , \qquad \quad \,\,\,\, \quad  V_{2N}={\theta[^0_0]^N-\theta[^0_1]^N\over 2\eta^N}\, ,\nonumber \\
&&\displaystyle \,S_{2N}\,={\theta[^1_0]^N+(-i)^N\theta[^1_1]^N\over 2\eta^N}\, , \quad \quad C_{2N}={\theta[^1_0]^N-(-i)^N\theta[^1_1]^N\over 2\eta^N}\, ,
\label{charac}
\end{eqnarray}
$Z_{\N=4}$ can be written in a more compact form,
\begin{equation}
Z_{\N=4}= O^{(0)}_{2,2}\,  O^{(1)}_{2,2}\, O^{(2)}_{2,2}\, O^{(3)}_{2,2} \left\{ V_8-S_8\right\}\left\{\bar O_{16}+\bar S_{16}\right\} \left\{ \bar O'_{16}+\bar S'_{16}\right\},
\end{equation}
where the $E_8$ characters are expressed as sums of $SO(16)$ ones, $\bar O_{16}+\bar S_{16}$.

Next, we define a cousin $\N=4$ model, obtained by implementing a $\Z_2$-shift along the direction $X^4$ that is correlated to the characters of both $SO(16)$ factors. This is done  by inserting in the partition function $Z_{\N=4}$ the sign $S_R=e^{i\pi[g^1_1(\gamma+\gamma')+h^1_1(\delta+\delta')]}$, whose effect is to break $E_8\times E_8'\to SO(16)\times SO(16)'$. 

Finally, we include the non-trivial SSS sign $S_L=e^{i\pi(g^1_1a+h^1_1b+h^1_1g^1_1)}$, in order to break spontaneously $\N=4\to \N=0$. In total, the non-susy partition function is obtained from Eq. (\ref{N4}) by replacing
\begin{equation}
\Gamma_{2,2}(T_1,U_1)\to {1\over 2}\sum_{h^1_1,g^1_1}\Gamma_{2,2}\Big[{}^{h_1^1,\, 0} _{g_1^1,\, 0} \Big]\!(T_1,U_1)\;  e^{i\pi[g^1_1(a+\gamma+\gamma')+h^1_1(b+\delta+\delta')+h^1_1g^1_1]}\,.
\end{equation}
Defining new ``shifted characters'' associated to the $1^{\rm st}$ 2-torus as  
 \begin{equation}
 \label{defO}
O^{(1)}_{2,2}[^{h}_{g}]={\Gamma_{2,2}[^{h,\, 0}_{0,\, 0}](T_1,U_1)+(-)^g\, \Gamma_{2,2}[^{h,\, 0}_{1,\, 0}](T_1,U_1)\over 2\eta^2\bar\eta^2}\, ,
 \end{equation}
 the final partition function we consider is \cite{snc}
\begin{eqnarray}
\label{Zsss}
Z^{\rm sss}_{\N=0}=O^{(0)}_{2,2} \, O^{(2)}_{2,2}\,  O^{(3)}_{2,2}\!\!\!\!\!\!\!\!\!\!\!&\Big[\; \;\,O_{2,2}^{(1)}[^0_0] \left\{ V_8(\bar O_{16}  \bar O'_{16}+ \bar S_{16} \bar S'_{16} ) -S_8 (\bar O_{16}\bar S'_{16} + \bar S_{16} \bar O'_{16})\right\} \nonumber \\
& \;\;\;\;\;\,+O_{2,2}^{(1)}[^0_1]\left\{ V_8 (\bar O_{16}\bar S'_{16} + \bar S_{16} \bar O'_{16}) -S_8  (\bar O_{16} \bar O'_{16}+ \bar S_{16}\bar S'_{16} ) \right\}\phantom {\, \Big]\,.}\nonumber\\
&\;\;\;\;\;+O_{2,2}^{(1)}[^1_0]\left\{O_{8} (\bar V_{16} \bar C'_{16}+ \bar C_{16}\bar V'_{16} ) -C_8 (\bar V_{16}\bar V'_{16} + \bar C_{16}\bar C'_{16} )\right\} \phantom {\, \Big]\,.}\nonumber\\
& \;\;\;\;\;+O_{2,2}^{(1)}[^1_1]\left\{O_8(\bar V_{16}\bar V'_{16} + \bar C_{16}\bar C'_{16}) -C_8 (\bar V_{16} \bar C'_{16}+ \bar C_{16}\bar  V'_{16} )\right\}\, \Big]\,.
 \end{eqnarray}
 Because the sign $S_LS_R$ couples the $\Z_2$-shift to the left- and right-moving spinorial representations, namely the spacetime fermions and the spinorial representations of both $SO(16)$ factors, we shall refer to this particular breaking as the ``spin-spin-spin'' (sss) breaking.  The latter can be compared to the breaking of the initial $\N=4$, $E_8\times E_8'$ theory using the SSS sign $S_L$ only, referred as ``spin''~(s) breaking, which leads to the partition function 
  \begin{equation}
Z^{\rm s}_{\N=0}= O^{(0)}_{2,2} \, O^{(2)}_{2,2} \, O^{(3)}_{2,2}
\left\{ O^{(1)}_{2,2}[^0_0]V_8 - O^{(1)}_{22}[^0_1]S_8 - O^{(1)}_{22}[^1_0]C_8 + O^{(1)}_{22}[^1_1]O_8\right\}
\left\{\bar O_{16}+\bar S_{16}\right\} \left\{ \bar O'_{16}+\bar S'_{16}\right\}.
\end{equation}
In this case, the shift is coupled to the spacetime fermions only, as  in the models at finite temperature, and the purely right-moving conformal blocks remain factorized.

To discuss the spectra, we use the explicit form of the $O^{(1)}_{2,2}[^{h}_{g}]$ character,
\begin{equation}
O^{(1)}_{2,2}[^h_g]={1\over (\eta\bar \eta)^2}\sum_{p_L,p_R}q^{{1\over 2}\abs p_L\abs^2}\, \bar q^{{1\over 2}\abs p_R\abs^2} ,
\end{equation}
where the sum runs overs the four integers $k^1,n^1, m^2,n^2$ appearing in the definitions
\begin{eqnarray}
&&\displaystyle p_L={1\over \sqrt{2\, \Im T_1\, \Im U_1}}\left[U_1(2k^1+g)-m^2+{T_1\over 2}\big(2n^1+h\big)+T_1U_1 n^2\right]\nonumber\\
&&\displaystyle p_R={1\over \sqrt{2\, \Im T_1\, \Im U_1}}\left[\bar U_1(2k^1+g)-m^2+{T_1\over 2}\big(2n^1+h\big)+T_1\bar U_1 n^2\right].
\end{eqnarray}
Note that the s-breaking contains an $O_8\bar O_{16}\bar O'_{16}$ sector, which leads to 2 tachyonic scalars, when ${1\over 2}\abs p_L\abs ^2-{1\over 2}={1\over 2}\abs p_R\abs^2-1<0$. On the contrary, with the sss-breaking, the dangerous left-moving character $O_8$ is accompanied by a right-moving sector that starts at the massless level, $\bar V_{16}\bar V'_{16}$. Therefore, contrary to the s-case, the sss-model is free of any Hagedorn-like instability. 

We are mostly interested in the light spectrum of the sss-model, when the gravitino mass $m_{3/2}$ given in Eq. (\ref{m32}) is much lower than the string scale, \ie when $\Im T_1\gg 1$ and $U_1=\O(1)$. Since all states in the sectors $O_{2,2}^{(1)}[^1_g]$ have non-trivial winding numbers along the large compact direction $X^4$, they are supermassive. Thus, we only have to analyze the sectors $O_{2,2}^{(1)}[^0_g]$.\\

\noindent {\large \em Sector $O^{(1)}_{2,2}[^0_0]$}

Massless bosons are present in the sector $O^{(0)}_{2,2} \, O^{(1)}_{2,2}[^0_0] \, O^{(2)}_{2,2}\, O^{(3)}_{2,2}\,   V_8 \bar O_{16} \bar O'_{16}$. They are the components of the graviton, antisymmetric tensor and dilaton, together with the gauge bosons of a gauge group $G=G^{(1)}\times G^{(2)}\times G^{(3)}\times SO(16)\times SO(16)'$, where $G^{(I)}$ arises from the lattice of the $I^{\rm th}$ $\mbox{2-torus}$. We have $G^{(1)}=U(1)^2$, but  $G^{(I)}$, $I=2,3$, can be any group of rank 2. It is generically $U(1)^2$ but can be enhanced to $SU(2)\times U(1)$, $SU(2)^2$ or $SU(3)$, when $(T_I,U_I)$ sits at particular points. The counting of states is as follows,
\begin{eqnarray}
d({\rm Bosons}[^0_0])\!\!\!\!\!&&= d(V_8)\!\left[d(O^{(0)}_{2,2})+d(O^{(1)}_{2,2}[^0_0])+d(O^{(2)}_{2,2}) + d(O^{(3)}_{2,2})+d(\bar O_{16})+d(\bar O'_{16})\right]\nonumber \\
&&=8\times \big[2+2+d(G^{(2)}) + d(G^{(3)}) +8\times 15+ 8\times 15  \big] \nonumber \\
&&=\underline{8\times \big[244+d(G^{(2)})+d(G^{(3)})\big]} .
\end{eqnarray}

Massless fermions are present in the sectors $-O^{(0)}_{2,2} \, O^{(1)}_{2,2}[^0_0] \, O^{(2)}_{2,2}\, O^{(3)}_{2,2}\,   S_8 (\bar O_{16} \bar S'_{16}+\bar S_{16} \bar O'_{16})$. They are  in the spinorial representations of $SO(16)$ or $SO(16)^{\prime}$, and their degeneracy is
\begin{equation}
d({\rm Fermions}[^0_0])= d(S_8)\!\left[d(\bar S'_{16})+d(\bar S_{16})\right]=\underline{8\times 256}\, .
\end{equation}

The above bosonic and fermionic degrees of freedom are accompanied by light towers of pure KK states arising from the $1^{\rm st}$ 2-torus. Their momenta along the large directions $X^4$ and $X^5$ are $2k^1$ and $m^2$, and their masses are of order $m_{3/2}$. 
\\

\noindent {\large \em  Sector $O^{(1)}_{2,2}[^0_1]$}

Light towers of pure KK bosonic states arise in the sectors $O^{(0)}_{2,2} \, O^{(1)}_{2,2}[^0_1] \, O^{(2)}_{2,2}\, O^{(3)}_{2,2}\,   V_8 (\bar O_{16} \bar S'_{16}+\bar S_{16} \bar O'_{16})$. Their momenta along $X^4$ and $X^5$ are $2k+1$ and $m^2$, and their multiplicity is 
 \begin{equation}
d({\rm Bosons}[^0_1])= d(V_8)\!\left[d(\bar S'_{16})+d(\bar S_{16})\right] =\underline{8\times 256}\, .
\end{equation} 

Light towers of pure KK fermionic states arise in the sector $-O^{(0)}_{2,2} \, O^{(1)}_{2,2}[^0_1] \, O^{(2)}_{2,2}\, O^{(3)}_{2,2}\,  S_8 \bar O_{16} \bar O'_{16}$, with momenta $2k+1$ and $m^2$. Their counting is 
\begin{eqnarray}
d({\rm Fermions}[^0_1])\!\!\!\!\!&&= d(S_8)\!\left[d(O^{(0)}_{2,2})+d(O^{(1)}_{2,2}[^0_1])+d(O^{(2)}_{2,2}) + d(O^{(3)}_{2,2})+d(\bar O_{16})+d(\bar O'_{16})\right]\nonumber \\
&&=8\times \big[2+2+d(G^{(2)}) + d(G^{(3)}) +8\times 15+ 8\times 15  \big] \nonumber \\
&&=\underline{8\times \big[244+d(G^{(2)})+d(G^{(3)})\big]} .
\end{eqnarray}

In total, the degeneracies encountered in the sectors $O_{2,2}^{(1)}[^0_g]$ yield
\begin{equation}
n_{\rm F}=8\times 256\; , \quad \qquad n_{\rm B}=8\times \big[244+d(G^{(2)})+d(G^{(3)})\big],
\end{equation} 
which are equal when $d(G^{(2)})+d(G^{(3)})=12$, \ie for\footnote{If one considers a compactification on $T^2\times T^4$, another group $G_2\times U(1)^2$ is also possible, where the exceptional Lie group $G_2$ is realized at some special point of the  $\Gamma_{4,4}$-lattice. This case yields $\N=4$ and $\N=2$ super no-scale models.}
\begin{equation}
G^{(2)}\times G^{(3)}=SU(2)^4\qquad \mbox{or} \qquad G^{(2)}\times G^{(3)}=SU(3)\times SU(2)\times U(1)\, .
\end{equation}
Modulo T-duality, the solution $SU(2)^4$ is obtained at the self-dual point  $T_2=U_2=T_3=U_3=i$. Moreover, locally around this point, some of the $SU(2)$ factors are spontaneously broken to $U(1)$, so that $n_{\rm B}$ decreases and $\V_{\mbox{\scriptsize 1-loop}}$ given in Eq. (\ref{v1}) becomes positive. Therefore, the 1-loop effective potential is positive semi-definite at the above self-dual point, with flat directions  $m_{3/2}$ and $U_1$. The model is thus attracted dynamically to a point in moduli space characterized by a super no-scale structure. 

The second solution, $SU(3)\times SU(2)\times U(1)$, is realized modulo T-duality at $T_2=U_2=e^{i\pi/3}$, $T_3=U_3$ arbitrary. Locally around this complex line, the gauge group $G$ is as before spontaneously broken to a subgroup and $n_{\rm B}$ decreases. Thus,  the 1-loop effective potential is locally positive semi-definite, with flat directions $m_{3/2}$, $U_1$ and $T_3=U_3$. Again, the model is naturally super no-scale; the time-dependent moduli trajectories being attracted to these points.  

Out of the super no-scale regime, when $m_{3/2}=\O(M_{\rm s})$, the effective potential does not vanish anymore. The generic massless states listed above are not accompanied by light KK modes, the latter having now masses of order of the string scale. However, states with non-trivial momentum and winding numbers along the $1^{\rm st}$ 2-torus, which are supermassive in the super no-scale regime, can be massless at special points in moduli space, when $T_1,U_1=\O(1)$ :
\\

\noindent {\large \em Sector $O^{(1)}_{2,2}[^0_0]$}

Additional massless bosons arise in the sector $O^{(0)}_{2,2} \, O^{(1)}_{2,2}[^0_0] \, O^{(2)}_{2,2}\, O^{(3)}_{2,2}\,   V_8 \bar O_{16} \bar O'_{16}$, when ${1\over 2}\abs p_L\abs ^2={1\over 2}\abs p_R\abs^2-1=0$. In the case the $1^{\rm st}$ 2-torus is factorized in two circles of radii $R_1$ and $R_2$, these conditions  are realized for momenta and winding numbers $m^2=-n^2=\pm1$, $k^1=n^1=0$, at the self-dual point $R_2=1$, $R_1$ arbitrary. They are two gauge bosons,
\begin{equation}
d(\mbox{Extra Bosons}[^0_0])= d(V_8)\times d(O^{(1)}_{2,2}[^0_0])=\underline{8\times 2}\, ,
\end{equation}
which enhance the gauge group factor associated to the $1^{\rm st}$ 2-torus, $G^{(1)}=U(1)\times SU(2)$. 
\\

\noindent {\large \em Sector $O^{(1)}_{2,2}[^1_1]$}

Extra  massless bosons arise in the sector $O^{(0)}_{2,2} \, O^{(1)}_{2,2}[^1_1] \, O^{(2)}_{2,2}\, O^{(3)}_{2,2}\, O_8 \bar V_{16} \bar V'_{16}$, when ${1\over 2}\abs p_L\abs^2-{1\over 2}={1\over 2}\abs p_R\abs^2=0$. For a factorized $1^{\rm st}$ 2-torus, this is realized for $2k^1+1=2n^1+1=\pm 1$, $m^2=n^2=0$ at the fermionic point $R_1=\sqrt{2}$, $R_2$ arbitrary. They are two scalars in the bi-vectorial representation of $SO(16)\times SO(16)'$, thus with multiplicity
 \begin{equation}
d(\mbox{Extra Bosons}[^1_1])= d(O^1_{2,2}[^1_1])\, d(\bar V_{16}) \, d(\bar V'_{16})=\underline{2\times 16\times 16} \, .  
\end{equation}


\section{Conclusion}
\label{cl}

In this work, we have considered $\N=1$ no-scale models in string theory, where the spontaneous breaking of supersymmetry to $\N=0$ is implemented at tree level by a SSS mechanism. Imposing the gravitino mass $m_{3/2}$ to be light, compared to the string scale, imposes a subspace of the internal space to be large. We have recalled that this often induces large quantum corrections to the gauge couplings, which implies the string coupling to be fine-tuned, a fact that is non-natural and known as the ``decompactification problem''. In the context of heterotic $\Z_2\times \Z_2$ orbifolds or moduli-deformed fermionic constructions, we have reviewed a solution to this puzzle that is valid, when some of the $\Z_2$ actions is freely acting \cite{N=0thresh}.   

We  also pointed out that the smallness of $m_{3/2}$ implies the  1-loop effective potential to be dominated by KK states, and to scale generically as $(n_{\rm F}-n_{\rm B})m_{3/2}^4$, where $n_{\rm F}$  and $n_{\rm B}$ count the numbers of massless fermions  and bosons. If this is interesting from the point of view of the gauge hierarchy problem, the order of magnitude of the cosmological term generated at 1-loop is still very large, compared to the presently observed one. Thus, we have defined ``super no-scale-models'' as being those for which the no-scale structure persists at 1-loop \cite{snc}. In other words, the gravitino mass is a flat direction of a positive semi-definite 1-loop effective potential. In practice, we have found string models that satisfy $n_{\rm F}=n_{\rm B}$, which yields an exponentially suppressed vacuum energy at 1-loop for arbitrary $m_{3/2}$, as long as the latter is  small compared to the string scale. In these examples, the moduli that do not participate in the susy breaking are naturally and dynamically attracted to points characterized by the super no-scale structure. 

Even if this is not {\em a priori}  required, the examples of super no-scale models we have pointed out are in the class of $\N=1\to \N=0$, $\Z_2\times \Z_2$ theories, where a $\Z_2$ is freely acting, thus avoiding the decompactification problem. However, we have stressed that a $\Z_2$ free action implies the massless spectrum to be non-chiral. Therefore, starting with semi-realistic $\N=1$ chiral models, it it still challenging to find a suitable way to implement a spontaneous $\N=1\to \N=0$ breaking that satisfies : chirality of the $\N=0$ spectrum, naturalness (no fine-tuning) and analytic control, to preserve the full predictability of the string model. 

Finally, we mention that other chiral or non-chiral, non-supersymmetric models have been considered in Ref. \cite{ADM}, including cases where $n_{\rm F}=n_{\rm B}$. However, they generically suffer from the decompactification problem and the eventual presence of instabilities at finite points in moduli space.  

  
 \section*{Acknowledgement}
 
We are grateful to C. Angelantonj, C. Bachas, G. Dall'Agata, A. Faraggi, I. Florakis and  J. Rizos  for fruitful discussions. 
The work of C.K. is partially supported by his  Gay Lussac-Humboldt Research Award 2014, in the  Ludwig Maximilians University and Max-Planck-Institute for Physics. H.P. would like to thank the Laboratoire de Physique Th\'eorique of Ecole Normale Sup\'erieure for hospitality. 


\end{document}